\documentclass[a4paper,notitlepage]{article} 

\usepackage[dvips]{graphicx}
\usepackage{amsmath}
\usepackage{amsfonts}
\usepackage{amssymb}

\newcommand{\nwc}{\newcommand}
\nwc{\ket}[1]{|#1\rangle}
\nwc{\bra}[1]{\langle#1|}
\nwc{\scal}[2]{\bra{#1}#2\rangle}
\nwc{\be}{\begin{equation}}
\nwc{\ee}{\end{equation}}
\nwc{\bea}{\begin{eqnarray}}
\nwc{\eea}{\end{eqnarray}}

\def\O{\Omega}

\def\la{\lambda}
\def\l{\left}
\def\r{\right}
\def\Neel{N{\'e}el\ }
\nwc{\bb}{\boldsymbol{\beta}}
\nwc{\ba}{\boldsymbol{\alpha}}
\nwc{\cA}{{\textsf{A}}}
\nwc{\cD}{{\textsf{D}}}
\nwc{\cQ}{{\textsf{Q}}}
\nwc{\cR}{{\textsf{R}}}
\nwc{\cS}{{\textsf{S}}}
\nwc{\cT}{{\textsf{T}}}
\nwc{\cV}{{\textsf{V}}}
\nwc{\Sv}{{\mathbf{S}}}
\nwc{\Hv}{{\mathbf{H}}}
\nwc{\ov}{\widehat{ {\mbox{\boldmath ${\O}$}} }}
\nwc{\fiv}{\widehat{ {\mbox{\boldmath $\phi$}}  }}
\nwc{\nv}{\widehat{ {\mbox{\boldmath $n$}}  }}
\nwc{\zv}{\widehat{ {\mbox{\boldmath $z$}}  }}
\nwc{\tS}{\widetilde{S}}
\nwc{\cH}{{\cal H}}
\nwc{\cG}{{\cal G}}

\begin{document}


\title{\bf LOW-DIMENSIONAL SPIN SYSTEMS: HIDDEN SYMMETRIES, CONFORMAL FIELD
THEORIES AND NUMERICAL CHECKS}

\author{
C. Degli Esposti Boschi$^1$, E. Ercolessi$^{1,2}$ and G. Morandi$^{1,2}$ \\ \\
{\small \it $^1$Unit{\`a} di ricerca INFM di Bologna,}\\ 
{\small \it $^2$Dipartimento di Fisica, Universit{\`a} di Bologna and INFN,} \\
{\small \it viale Berti-Pichat, 6/2, I-40127, Bologna, Italia}
}

\maketitle

{\small
\begin{center}
{\bf Abstract:}
\end{center}
We review here some general properties of antiferromagnetic Heisenberg spin
chains, emphasizing and discussing the r\^{o}le of hidden symmetries in the
classification of the various phases of the models. We present also some
recent results that have been obtained with a combined use of Conformal Field
Theory and of numerical Density Matrix Renormalization Group techniques.
}


\section{Introduction and Summary.}

For quite some time low-dimensional magnetic systems (i.e. (quantum) spins on
$1D$ and/or $2D$ lattices) have been considered essentially only as
interesting models in Statistical Mechanics with no realistic counterpart. It
is only in recent times that systems that can be considered to a high degree
of accuracy as assemblies of isolated or almost isolated spin chains and/or of
spin ladders (a few chains coupled together) have began to be produced and
have hence become experimentally accessible, thus renewing the interest in
their study, which is by now one of the most active fields of experimental and
theoretical research in Condensed Matter Physics.

In this paper we will discuss only some relevant properties of isolated spin
chains, referring to the literature \cite{DaR} for a general review of the
properties of spin ladders.

More than one decade ago it was pointed out \cite{DR,KT} that
\ \textit{integer} spin chains (more specifically, spin-$1$ chains, but
extensions to different values of the spin have also been devised in the
literature \cite{QLSC}) possess unexpected and highly nontrivial hidden
symmetries, whose spontaneous breaking manifests itself through the appearance
of unusual and highly nonlocal \ "string" order parameters. The string order
parameters, together with the more conventional magnetic order parameters, can
be used to classify the various phases that the phase diagram of
one-dimensional magnets can display.

In the present paper, which is a slightly enlarged version of the talk
presented by one of us (G.M.) at the $XIII-th$ Conference on
"\textit{Symmetries in Physics}"\footnote{The Conference, organized by Prof.
B. Gruber, was held in Schloss Mehrerau in Bregenz (Voralberg, Austria), \ in
the period 21-24 July, 2003.}, we will concentrate, without pretensions to
full generality, on the discussion of a few models of antiferromagnetic
Heisenberg chains, of their phase diagrams and on the r\^{o}le of hidden
symmetries in their explanation. The paper is organized as follows. In
Sect.$2$ we review some general facts concerning Heisenberg spin chains and
discuss how in the continuum limit one can map a "standard" (see below for the
terminology) Heisenberg chain onto an effective field theory described by a
nonlinear sigma-model, and how the presence in the latter of a topological
term can account for the radically different behaviors of integer versus
half-odd-integer spin chains. In Sect.$3$, concentrating on spin-$1$ chains,
we consider the effects of the addition to the "standard" model of biquadratic
exchange terms and/or of Ising-like as well as of single-ion anisotropies, and
how the addition of such terms can drive the model away from what is commonly
called the "Haldane phase" (again, see below for an explanation) towards other
phases. In this context we will introduce in a more explicit manner the notion
of hidden symmetries and discuss their r\^{o}le. Sects.$4$ and $5$ will be
devoted to the discussion of more recent results that have been obtained by
some of us \cite{DER} with a careful and combined use of analytical (effective
actions and Conformal Field Theory) and numerical (Density Matrix
Renormalization Group) techniques. The final Sect.$6$ is devoted to the
conclusions and to some general comments.

\section{General Features of Spin Chains.}

Let us begin by discussing here what can be considered \ as the "standard"
model of an isotropic antiferromagnetic ($AFM$) Heisenberg chain with
nearest-neighbor ($nn$) interactions, which is described by the Hamiltonian:%
\begin{equation}
\mathcal{H}=J%
{\displaystyle\sum\limits_{i=1}^{N}}
\overrightarrow{S}_{i}{\cdot}\overrightarrow{S}_{i+1}\equiv\mathcal{H}_{\bot
}+\mathcal{H}_{z}%
\end{equation}%
\begin{equation}
\mathcal{H}_{\bot}=\frac{1}{2}J_{\bot}%
{\displaystyle\sum\limits_{i=1}^{N}}
\left\{  S_{i}^{+}S_{i+1}^{-}+S_{i}^{-}S_{i+1}^{+}\right\}  ;\;\;
\mathcal{H}_{z}=J_{z}%
{\displaystyle\sum\limits_{i=1}^{N}}
S_{i}^{z}S_{i+1}^{z};\;J_{\bot}=J_{z}=J
\end{equation}
where, for each $i=1,...,N$, $\overrightarrow{S}_{i}$\ is a spin
operator\footnote{and: $S_{i}^{\pm}=S_{i}^{x}{\pm} iS_{i}^{y}$.}:%
\begin{equation}
\left[  S_{i}^{\alpha},S_{j}^{\beta}\right]  =i\hbar\delta_{ij}\varepsilon
^{\alpha\beta\gamma}S_{i}^{\gamma}; \; \alpha,\beta,\gamma
=x,y,z;\;\;\;\overrightarrow{S}_{i}^{2}=\hbar^{2}S(S+1)
\end{equation}
($S$ integer or half-odd integer) located \ at the $i-th$ site of a
one-dimensional lattice of $N$ sites, interacting with its neighbors with an
$AFM$ ($J>0$) $nn$ interaction of strength $J$. Later on we will consider more
general models in which $J_{\bot}\neq J_{z}$ will be allowed\footnote{$J_{\bot
}=0$, in particular, corresponds to the one-dimensional Ising model, a
trivially soluble \textit{classical } model. Notice however that an Ising
model in a \textit{transverse} magnetic field becomes a genuinely quantum and
nontrivial model.}.

It may be useful to define a vector $\overrightarrow{n}_{i}$ as:
$\overrightarrow{n}_{i}=:\overrightarrow{S}_{i}/\hbar S$, whereby:%
\begin{equation}
\left[  n_{i}^{\alpha},n_{j}^{\beta}\right]  =\frac{i}{S}\varepsilon
^{\alpha\beta\gamma}n_{i}^{\gamma};\;\overrightarrow{n}^{2}=1+\frac
{1}{S}%
\end{equation}

Although one is ultimately interested in the thermodynamic ($N\to
\infty$) limit, for finite $N$ one can adopt either periodic boundary
conditions ($PBC$'s), by imposing:%
\begin{equation}
\overrightarrow{S}_{i+N}=\overrightarrow{S}_{i}\;\forall i
\end{equation}
by which the system is actually considered to "live" on a circle, or open
boundary conditions ($OBC$'s), where $\overrightarrow{S}_{1}$ is coupled only
to $\overrightarrow{S}_{2}$ and $\overrightarrow{S}_{N}$ only to
$\overrightarrow{S}_{N-1}$\footnote{In which case the Hamiltonian should be
actually rewritten as: $\mathcal{H}=J%
{\displaystyle\sum\limits_{i=1}^{N-1}}
\overrightarrow{S}_{i}{\cdot}\overrightarrow{S}_{i+1}$.}. The Hamiltonian of
Eq.$(1)$ has an obvious (global) $O(3)$ symmetry and, for $PBC$'s, it is also
invariant under the (discrete) translation group of the lattice.

In the classical limit ($\hbar\to0$ \ and $S\to\infty$ \ with:
$\hbar S=const.$) the spins (the $\overrightarrow{n}_{i}$'s) become (see
Eq.$(4)$) classical vectors \ (and $\ \overrightarrow{n}_{i}\in {\mathbb S}^{2}$, the
unit sphere in $\mathbb{R}^{3}$). The minimum-energy configuration of the
spins corresponds to: $\overrightarrow{n}_{i}{\cdot}\overrightarrow{n}%
_{i+1}=const.=-1$. Neighboring spins are then aligned antiparallel to each
other and, in the absence of any external magnetic field, can point in a
common but otherwise arbitrary direction on the sphere. This is the
\textit{N\'{e}el state}. Let us remark that, at variance with the
\textit{ferro}magnetic ($J<0$) case, in which neighboring spins are all aligned
parallel, at the quantum level the N\'{e}el state is \textit{not} an
eigenstate of the Hamiltonian $(1)$. This points to the fact \ that
\ \textit{quantum} fluctuations will play a much more relevant r\^{o}le in the
(quantum) antiferromagnetic case than in the ferromagnetic one.

The classical energy of the N\'{e}el state is of course: $E_{N}=-JN(\hbar
S)^{2}$. In this state the $O(3)$ symmetry is spontaneously broken down to
$O(2)$\footnote{Translational symmetry, if present is also broken, as the
N\'{e}el state is not invariant under translations of a lattice spacing as the
original Hamiltonian but only of \textit{twice} the lattice spacing. This has
important consequences on the location of the Goldstone mode \cite{Am,Mor,Wa}
in momentum space, that we will not discuss here, however.}, and the state
exhibits \textit{long-range order} ($LRO$).

Elementary excitations over the N\'{e}el state are well-known to be in the form
of \textit{spin waves} \cite{Ma}: coherent deviations of the spins with a
dispersion: $\omega(\overrightarrow{k})\propto k$ in the long-wavelength
limit ($ka\ll1$, with $a$ the lattice spacing). Hence, the (classical) spectrum
of the Hamiltonian $(1)$ is \textit{gapless. }We would like to stress that
nothing of what has been said hitherto depends on the value of the spin.
\textit{At the classical level, the spin }$S$ \footnote{Or better $\hbar S.$%
}\textit{\ can be simply reabsorbed into a redefinition of the coupling
constant (}$J\to J(\hbar S)^{2}$) and will contribute only an
essentially irrelevant and additional multiplicative overall scale factor.

All this is elementary and well known. Let us turn now to the quantum
case\footnote{From now on we will set for simplicity $\hbar=1$.}. In the early
$30$'s Bethe \cite{Be} and Hulth\'{e}n \cite{Hu}, employing what has been known
since as the "Bethe-Ansatz", were able to show that the quantum $S=1/2$
\ Heisenberg chain is actually an \textit{integrable} model. We will not
discuss here the Bethe-Ansatz in any detail \cite{Ma}, but will only summarize
the main features of the solution of the $S=1/2$ model. The (exact) ground
state is nondegenerate, it exhibits only \textit{short-range }$AFM$
\ correlations, but \textit{no} $LRO$. Parenthetically, this is in agreement
with a general, and later, theorem \cite{Co}. The (staggered) static spin-spin
correlation functions:%
\begin{equation}
\mathcal{G}^{\alpha}(i-j)=(-1)^{\left\vert i-j\right\vert }\left\langle
S_{i}^{\alpha}S_{j}^{\alpha}\right\rangle ;\;\alpha=x,y,z
\end{equation}
where $\left\langle ...\right\rangle $ stands for the expectation value in the
ground state, are all equal and decay \textit{algebraically} to zero at large
distances. We recall here that genuine $LRO$ would imply (we omit here the
index $\alpha$):%
\begin{equation}
\underset{\left\vert i-j\right\vert \to\infty}{\lim}\mathcal{G}(i-j)=\mathcal{O}_{N}\neq0
\end{equation}
this defining the \textit{N\'{e}el order parameter} $\mathcal{O}_{N}$
\ (actually the square of the equilibrium staggered (i.e sublattice)
magnetization). On the other extreme, an \textit{exponential} decay of the
correlations of the form, say:%
\begin{equation}
\mathcal{G}(i-j)\underset{\left\vert i-j\right\vert \to\infty}%
{\approx}\exp\{-\left\vert i-j\right\vert /\xi\}P(\left\vert i-j\right\vert )
\end{equation}
with $P(.)$ some inverse power of \ $\left\vert i-j\right\vert $ would imply a
finite \textit{correlation length} $\xi$ and a mass gap (or, better, a spin
gap) $\Delta$ in the excitation spectrum roughly given by: $\Delta\propto
c\xi^{-1}$,with $c$ a typical spin-wave velocity. Algebraic decay of
correlations (formally corresponding to $\xi\to\infty$) implies then
that the system is \textit{gapless}. Summarizing, the main features of the
$S=1/2$ \ Heisenberg $AFM$ chain are that it has a (quantum) disordered ground
state, with only short-range $AFM$ correlations, and that it is gapless. It is
therefore a (actually the first) prototype of a (quantum disordered and)
\textit{quantum critical} system \cite{Sa}. It can be said then that, as
compared with the classical limit, the system remains gapless but
\textit{quantum fluctuations destroy }$LRO$.

About thirty years later Lieb, Schutz and Mattis \cite{LSM} ($LSM$) proved an
important theorem stating that an $S=1/2$ chain has either a degenerate ground
state or is gapless. No surprise that the Bethe solution obeys the
Lieb-Schutz-Mattis theorem, which is however of much wider reach, as it can
cover models that are more general than the "standard" $nn$ chain, such as,
e.g., the Majumdar-Ghosh \cite{MG} model, another integrable model that we will
nor discuss here, though. The results of $LSM$ were extended later on by other
authors \cite{A1} beyond $S=1/2$ to cover all the half-odd-integer values of
the spin. One can then take as rigorously proven that (at $T=0$)
\textit{isotropic half-odd-integer Heisenberg chains }(with constant $nn$ interactions)
\textit{are all quantum disordered and quantum critical} (i.e. gapless). This
result was thought for quite some time to be "generic", i.e. valid for chains
of arbitrary spin until, in the early $80$'s, Haldane \cite{Ha} put forward
what has become known since as \ "\textit{Haldane's conjecture}", according to
which half-odd-integer spin chains should be quantum disordered and gapless
but \textit{integer} spin chains should instead exhibit a spin gap and an
exponential decay of correlations. This implied that, contrary to what happens
in the classical limit, the physical behavior of spin chains should be a
\textit{highly discontinuous} function of the value of the spin.

Completely rigorous proofs of (the second part of) Haldane's conjecture are
still lacking. However, strong support to it comes from the analysis of the
continuum limit of the Heisenberg chain, which we will briefly describe now,
referring to the existing literature \cite{A2,Au,Fr} for more details.

The canonical partition function for the Hamiltonian of Eq.$\left(  1\right)
$ at temperature $\ T=(k_{B}\beta)^{-1}$ (with $k_{B}$ the Boltzmann
constant):%
\begin{equation}
\mathfrak{Z}=Tr\left\{  \exp\left[  -\beta\mathcal{H}\right]  \right\}
\end{equation}
can be written as a spin coherent-state path-integral \cite{KS}, whereby the
spin variables get replaced, inside the path-integral, by classical variables
according to:%
\begin{equation}
\overrightarrow{S}_{i}\to S\widehat{\Omega}_{i}%
\end{equation}
with $\widehat{\Omega}_{i}$ a classical unit vector: $\left\vert
\widehat{\Omega}_{i}\right\vert =1$. The next (and perhaps the most important)
step in Haldane's analysis is the parametrization of \ the $\widehat{\Omega
}_{i}$'s as\footnote{This is what is known in the literature as
"\textit{Haldane's mapping"}.}:%
\begin{equation}
\widehat{\Omega}_{i}=\left(  -1\right)  ^{i}\widehat{n}_{i}\sqrt{1-\left(
\frac{\overrightarrow{l}_{i}}{S}\right)  ^{2}}+\frac{\overrightarrow{l}_{i}%
}{S}%
\end{equation}
with: $\left\vert \widehat{n}_{i}\right\vert =1$ \ and: $\widehat{n}_{i}%
{\cdot}\overrightarrow{l}_{i}=0$. The $\widehat{n}_{i}$'s are assumed to be
slowly-varying (on the scale of the lattice spacing). In this way,
capitalizing, so-to-speak, on the information gained from the Bethe-Ansatz
solution of the $S=1/2$ model, they incorporate the information that the
system still retains some short-range $AFM$ ordering, which would be global
only for $\widehat{n}_{i}=const.$ (and $\overrightarrow{l}_{i}=0$). The
$\overrightarrow{l}_{i}$'s can be shown \cite{A2} to be the (local) generators
of angular momentum. In the semiclassical (large $S$) limit, an expansion of
the action in the path-integral up to lowest (second) order in the
$\overrightarrow{l}_{i}$'s is justified. Taking then the continuum limit
together with a gradient expansion, and integrating out the $\overrightarrow
{l}_{i}$'s, one ends up with the following expression for the partition
function:%
\begin{equation}
\mathfrak{Z}=\int\left[  \mathcal{D}\widehat{n}\right]  \delta\left(
\widehat{n}^{2}-1\right)  \exp\{-S_{E}-iS_{B}\}
\end{equation}
where $\left[  \mathcal{D}\widehat{n}\right]  $ stands for the functional
measure and the $\delta$ inside the integral is a functional $\delta$. The
first term in the action is given by:%
\begin{equation}
S_{E}=%
{\displaystyle\int\limits_{0}^{L}}
dx%
{\displaystyle\int\limits_{0}^{\beta}}
d\tau\left\{  \frac{1}{2g}\left[  \frac{1}{c}|\partial_{\tau}\widehat{n}%
|^{2}+c\left\vert \partial_{x}\widehat{n}\right\vert ^{2}\right]  \right\}
;\;\widehat{n}=\widehat{n}(x,\tau)
\end{equation}
where $L(=N \times {\rm lattice \; spacing})$ is the length of the chain, $g=2/S$ is the coupling constant and:
$c=2JS$ is the spin-wave velocity. This is simply the Euclidean action of an
$O(3)$ nonlinear sigma model \cite{Au,Fr,Zak} ($NL\sigma M$). The second term
is the integral of a Berry phase \cite{Sha}, and is given by:%
\begin{equation}
S_{B}=\frac{\theta}{4\pi}%
{\displaystyle\int\limits_{0}^{L}}
dx%
{\displaystyle\int\limits_{0}^{\beta}}
d\tau\widehat{n}{\cdot}\left(  \partial_{\tau}\widehat{n}{\times}\partial
_{x}\widehat{n}\right)
\end{equation}
with: $\theta=2\pi S$. The coefficient of $\theta$ is easily recognized to be
the Pontrjagin index \cite{BoTu,Mor2,Raja}, or winding number, of the map:%
\begin{equation}
\widehat{n}:\;\mathbb{R}_{comp}^{2}\mapsto\mathbb{S}^{2}%
\end{equation}
from spacetime compactified to a sphere and the two-sphere where $\widehat{n}$
takes values, and it is an integer: $S_{B}$ is therefore a topological term,
and: $S_{B}=2\pi nS,$ $n\in\mathbb{Z}$. Therefore, $\exp\left\{
-iS_{B}\right\}  \equiv1$ for integer $S$ \ ($\theta=0$ $\ \operatorname{mod}%
\left(  2\pi\right)  $), but: $\exp\left\{  -iS_{B}\right\}  =\left(
-1\right)  ^{n}$ ($\theta=\pi$ $\operatorname{mod}\left(  2\pi\right)  $) if
the spin is half-odd integer. This will generate interference between the
different topological sectors, and it is the at the heart of the different
behaviors of the two types of chains.

The pure ($\theta=0$ in our case) $(1+1)$ $O(3)$ $NL\sigma M$ \ is a
completely integrable model \cite{Za}. It has a unique ground state, and the
excitation spectrum is exhausted by a degenerate triplet of \textit{massive}
excitations that are separated from the ground state by a finite gap. On the
contrary, the $\theta=\pi$ model was shown \cite{RS} to be \textit{gapless}.
Therefore, Haldane's conjecture is fully confirmed by the analysis of the
continuum limit of the Heisenberg model.

We would like only to mention in passing that quite a similar behavior occurs
in spin \textit{ladders} \cite{DaR,DEMPR,RGS}, namely even-legged ladders are
gapped, while odd-legged ladders are gapped for integer spin and gapless for
half-odd-integer spin. This "\textit{even-odd}" effect has been
shown \cite{DEMPR,Si} to have the same topological origin an in single chains.

How do these results compare with the gaplessness (irrespective of the value
of $S$) of the $S\to\infty$ classical limit? The answer resides in the
dependence of the spin gap on $S$. Already at the mean-field level, but more
accurately from large-$N$ expansions and/or renormalization group
analyses \cite{Po}, it turns out that the spin gap $\Delta$ behaves as:
$\Delta\propto\exp\left\{  -\pi S\right\}$ for large $S$ 
\footnote{$\Delta\propto\exp\left\{  -\pi n_{l}S\right\}$ 
for spin ladders \cite{DEMPR}, where $n_{l}$ is the number of legs of the ladder.}. 
Hence, integer-spin
chains become exponentially gapless for large $S$, and the classical limit is
recovered correctly.

\section{More general Models. Hidden Symmetries and String Order Parameters.}

In view of what has been said up to now, the second part of Haldane's
conjecture is by far the most intriguing part of it. Therefore \textit{integer
}spin chains are the most interesting ones, and we will concentrate from now
on on $S=1$ chains.

What has been called in the previous Section the "standard" $AFM$ Heisenberg
model is actually a member of at least two larger families of models that we
will illustrate briefly here. The first class of models, that we will call
"$\theta-$\textit{models", } includes a biquadratic term in the spins, and is
described, setting $J=1$, by the Hamiltonian:%
\begin{equation}
\mathcal{H}=%
{\displaystyle\sum\limits_{i=1}^{N}}
\left\{  \cos\theta\left(  \overrightarrow{S}_{i}{\cdot}\overrightarrow{S}%
_{i+1}\right)  +\sin\theta\left(  \overrightarrow{S}_{i}{\cdot}\overrightarrow
{S}_{i+1}\right)  ^{2}\right\}
\end{equation}
with $\theta=0$ corresponding of course to the "standard" model. Most of the
phase diagram has been obtained \cite{KT} numerically, except for the points
at $\theta={\pm}\pi/4$, that correspond to integrable models. The point $\
\theta=\pi/4$ is the Sutherland model \cite{Su}, while $\theta=-\pi/4$ is the
integrable model \cite{B,T} of Babujian and Takhatajan\footnote{This point is
also known familiarly as the "Armenian point".}. Both models are gapless,
while the entire region $-\pi/4<\theta<\pi/4$ is known (numerically, again)
to be \textit{gapfull}. This whole region has been called the
"\textit{Haldane phase}". It includes a particularly interesting point that
has been studied extensively by Affleck, Kennedy, Lieb and Tasaki \cite{AKLT}
($AKLT$), namely $\theta=\theta^{\ast}$, with: $\tan\theta^{\ast}=1/3$. The
corresponding Hamiltonian (omitting an irrelevant overall numerical factor)
is given by:%
\begin{equation}
\mathcal{H}_{AKLT}=%
{\displaystyle\sum\limits_{i=1}^{N}}
\left\{  \overrightarrow{S}_{i}{\cdot}\overrightarrow{S}_{i+1}+\frac{1}%
{3}\left(  \overrightarrow{S}_{i}{\cdot}\overrightarrow{S}_{i+1}\right)
^{2}\right\}
\end{equation}

This model is not completely integrable, but the ground state is known, it is
unique in the thermodynamic limit and can be exhibited explicitly. The
ultimate reason for this is that, apart from numerical constants, the $i-th$
term in curly brackets is just:%
\begin{equation}
\overrightarrow{S}_{i}{\cdot}\overrightarrow{S}_{i+1}+\frac{1}{3}\left(
\overrightarrow{S}_{i}{\cdot}\overrightarrow{S}_{i+1}\right)  ^{2}=2\left[
P_{2}(i,i+1)-\frac{1}{3}\right]
\end{equation}
where $P_{2}(i,i+1)$ is the projector \cite{Me} onto the state of total spin
$S_{tot}=2$ of the pair of $S=1$ spins located at sites $i$ and $i+1$.
Therefore, the ground state of $\ \mathcal{H}_{AKLT}$ must lie in the sector
of the Hilbert space that is annihilated by all the projectors. It was shown
by $AKLT$ that the exact ground state (also called the "Valence-Bond-Solid"
($VBS$) state) can be constructed as a linear superposition of states
$\Phi_{\sigma}$ that have the following characteristics. Let: $\sigma=\left\{
\sigma_{1},...\sigma_{N}\right\}  $ be a given spin configuration, with:
$\sigma_{i}=0,{\pm}1,$ $i=1,...,N$. Then, $\Phi_{\sigma}$ \ is such that:

$i)$ $S_{i}^{z}\Phi_{\sigma}=\sigma_{i}\Phi_{\sigma}$ and moreover: $ii)$ If a
given spin is, say, $+1$, then the next \textit{nozero} spin must be $-1$, and
viceversa. Typical such states correspond therefore to spin configurations of
the form:%
\begin{equation}
\sigma=\left\{ +00-0+-000+...\right\}
\end{equation}

In other words, "up" and "down" spins do alternate in $\Phi_{\sigma}$, but
their spatial distribution is completely \textit{random}, as an arbitrary
number of zeroes can be inserted between any two nonzero spins. So, if a given
spin in nonzero, we can predict \textit{what} the value of the next nonzero
spin will be, but not \textit{where} it will be located. There is therefore no
long-range (N\'{e}el) order in any conventional sense in the $VBS$ ground
state, but a sort of "\textit{Liquid N\'{e}el Order}" ($LNO$). Conventional
N\'{e}el order would be characterized by a nonvanishing of (at least one of)
the \textit{N\'{e}el order parameters}:%
\begin{equation}
\mathcal{O}_{N}^{\alpha}=\underset{|i-j|\to\infty}{\lim}\left(
-1\right)  ^{|i-j|}\left\langle S_{i}^{\alpha}S_{j}^{\alpha}\right\rangle
;\;\; \alpha=x,y,z
\label{NOPs}
\end{equation}
In the $VBS$ state and (numerically) in the whole of the Haldane phase one
finds instead \cite{AKLT,KT}: $\mathcal{O}_{N}^{\alpha}=0,\alpha=x,y,z$, and
this is consistent with the absence of a "rigid" N\'{e}el order.

There remains however what we have called the "liquid" N\'{e}el order, and it has
been argued convincingly in the literature \cite{DR,KT} that this is connected
with the nonvanishing of a novel class of order parameters that we will
discuss now briefly. Let us begin by defining the \textit{string correlation
functions} as:%
\begin{equation}
\mathcal{G}_{S}^{\alpha}(n)=:-\left\langle S_{0}^{\alpha}\exp\left[  i\pi%
{\displaystyle\sum\limits_{l=1}^{n-1}}
S_{l}^{\alpha}\right]  S_{n}^{\alpha}\right\rangle ;\;\;
\alpha=x,y,z;\;n>0
\label{SOcf}
\end{equation}
These are similar to the standard two-point correlation functions:%
\begin{equation}
\mathcal{G}^{\alpha}(n)=:\left(  -1\right)  ^{n}\left\langle S_{0}^{\alpha
}S_{n}^{\alpha}\right\rangle
\end{equation}
whose asymptotic ($n\to\infty$) limit yields the N\'{e}el order
parameter(s), except that a \textit{string} of exponentials of intermediate
spins has been inserted between the leftmost and the rightmost spins.

The \textit{string order parameters }($SOP$'s)\ $\mathcal{O}_{S}^{\alpha}$ are
then defined as:%
\begin{equation}
\mathcal{O}_{S}^{\alpha}=\underset{n\to\infty}{\lim}\mathcal{G}%
_{S}^{\alpha}(n)
\label{SOPs}
\end{equation}

It turns out \cite{AKLT,GA} that the string correlation functions are strictly
\textit{constant} in the $AKLT$ ground state, namely:%

\begin{equation}
\mathcal{G}_{S}^{\alpha}(n)\equiv const.=\mathcal{O}_{S}^{\alpha}=\frac{4}{9}%
\end{equation}

The ground-state spin-spin correlation functions have also been evaluated
exactly for the $VBS$ state \cite{AKLT}, and they turn out to be given by:%
\begin{equation}
\mathcal{G}^{\alpha}(n)=\frac{4}{3}\left(  \frac{1}{3}\right)  ^{n}%
\end{equation}
In other words: $\mathcal{G}^{\alpha}(n)\propto\exp\left\{  -n/\xi
_{AKLT}\right\}  $, where the \textit{correlation length} $\xi_{AKLT}$ is
given, in units of the lattice spacing, by:%
\begin{equation}
\xi_{AKLT}=\frac{1}{\log3}\simeq0.91
\end{equation}
less than unity in units of the lattice spacing, implying a rather large spin gap.

So far for the ground state of the $AKLT$ model. String and ordinary
correlation functions as well as N\'{e}el and string order parameters have
also been evaluated (numerically away from the $AKLT$ point) for other points
of the Haldane phase \cite{GA}. For example, at the Heisenberg point, exact
diagonalization methods\footnote{With the Lanczos method and for chains with
up to no more than $14$ sites.} have shown that the string correlation
functions are not strictly constant, but still decay exponentially to a value
of the string order parameter that is somewhat smaller ($\mathcal{O}%
_{S}^{\alpha}\simeq0.36...$) than the $AKLT$ value ($\mathcal{O}_{S}^{\alpha
}=4/9\simeq0.44...$) but still nonzero. The spin correlation length was also
found \cite{GA} to be slightly larger than the $AKLT$ value, but still finite.
So, there is convincing evidence that the entire Haldane phase is
characterized by \textit{vanishing N\'{e}el order parameters} but by
\textit{nonzero }$SOP$\textit{'s. }There is also convincing numerical
evidence \cite{GA} that the string order parameters vanish at the integrable
boundaries of the Haldane phase (i.e. for $\theta={\pm}\pi/4$).

That the nonvanishing of the $SOP$'s is connected to the breaking of a
symmetry, and hence to the onset of an ordering that is not apparent in the
original Hamiltonian was clarified in a seminal paper by Kennedy and
Tasaki \cite{KT} ($KT$). With reference to a given configuration $\left\{
\sigma\right\}  $, and defining $N(\sigma)$ as the number of odd sites at
which the spins are zero, one defines a new configuration $\left\{
\overline{\sigma}\right\}  $ via:%
\begin{equation}
\overline{\sigma}_{i}=\exp\left[  i\pi%
{\displaystyle\sum\limits_{j=1}^{i-1}}
\sigma_{j}\right]  \sigma_{i}%
\end{equation}
and then a unitary operator $U$ via:%
\begin{equation}
U\Phi_{\sigma}=\left(  -1\right)  ^{N(\sigma)}\Phi_{\overline{\sigma}}%
\end{equation}
In a nutshell, the action of $U$ amounts to leaving the first nonzero spin
unchanged and to flipping every other nonzero spin proceeding to the right of
the chain. For example:%
\begin{equation}%
\begin{array}
[c]{c}%
\left\{  ++0-+00+0-0++\right\}  \mapsto\left\{  +-0--00+0+0+-\right\} \\
\left\{  0+-00+00-+00-\right\}  \mapsto\left\{  0++00+00++00+\right\}
\end{array}
\end{equation}
and so on. It is obvious that $U$ is a unitary\footnote{Notice also that:
$N\left(  \sigma\right)  =N\left(  \overline{\sigma}\right)  $, as zero spins
are mapped into zero spins.}. What is less obvious is that the unitary
transformation is a \textit{nonlocal} one, in the sense that $U$ \ cannot be
written as a product of unitary operators acting at each single site. This has
the important consequence that symmetries that are local (in the above sense)
for the Hamiltonian $\mathcal{H}$ will of course remain symmetries of the
transformed Hamiltonian $\widetilde{\mathcal{H}}$ (as $U$ is unitary) but need
not \ survive as \textit{local} symmetries of $\widetilde{\mathcal{H}}$.
Specifically, the symmetry group of $\mathcal{H}$ is $SU(2)$, that includes a
discrete $Z_{2}{\times} Z_{2}$ subgroup of rotations of $\pi$ around the
coordinate axes. Explicitly, the transformed Hamiltonian has the
form \cite{KT}:%
\begin{equation}
\widetilde{\mathcal{H}}=%
{\displaystyle\sum\limits_{i}}
\left\{  \cos\theta h_{i}+\sin\theta\left(  h_{i}^{2}\right)  \right\}
\end{equation}
where:%
\begin{equation}
h_{i}=-S_{i}^{x}S_{i+1}^{x}+S_{i}^{y}\exp\left\{  i\pi\left(  S_{i}%
^{z}+S_{i+i}^{z}\right)  \right\}  S_{i+1}^{y}-S_{i}^{z}S_{i+1}^{z}%
\end{equation}
and it evident that $Z_{2}{\times} Z_{2}$ is the only \textit{local} surviving
symmetry group of the transformed Hamiltonian $\widetilde{\mathcal{H}}$. Even
more important is how the string order parameters transform. The result
is \cite{KT}:%
\begin{equation}
\mathcal{O}_{S}^{\alpha}\left(  \mathcal{H}\right)  =\mathcal{O}%
_{ferro}^{\alpha}\left(  \widetilde{\mathcal{H}}\right)
\end{equation}
where:%
\begin{equation}
\mathcal{O}_{ferro}^{\alpha}\left(  \widetilde{\mathcal{H}}\right)
=\underset{|i-j|\to\infty}{\lim}\left\langle S_{i}^{\alpha}%
S_{j}^{\alpha}\right\rangle |_{\widetilde{\mathcal{H}}}%
\end{equation}
and the r.h.s stands here for an average taken w.r.t. the ground state of the
transformed Hamiltonian. The transformed order parameter is now a
\textit{ferromagnetic} order parameter. Therefore: $\mathcal{O}_{S}^{\alpha
}\left(  \mathcal{H}\right)  \neq0\Longrightarrow\mathcal{O}_{ferro}^{\alpha
}\left(  \widetilde{\mathcal{H}}\right)  \neq0$, and this implies the onset of
a spontaneous ferromagnetic polarization in the $\alpha-th$ direction in the
ground state of $\widetilde{\mathcal{H}}$. This in turns entails a partial (if
$\mathcal{O}_{ferro}^{\alpha}\left(  \widetilde{\mathcal{H}}\right)  \neq0$
for just one value of $\alpha$) or total (if this happens in more than one
direction) spontaneous breaking of the discrete $Z_{2}{\times} Z_{2}$ symmetry.
It is known \cite{Am,Wa} that spontaneous breaking of a continuous symmetry is
accompanied by massless excitations (the Goldstone modes), while breaking of a
discrete symmetry usually implies the opening of a gap (the most conspicuous
and familiar example being the $2D$ Ising model). \ Therefore, $KT$ \ were led
to consider the spontaneous breaking of the $Z_{2}{\times} Z_{2}$ symmetry as
the origin of the Haldane gap.

One has however to be a bit careful on this point. It appears to be true that
spontaneous (partial or total) breaking of the $Z_{2}{\times} Z_{2}$ symmetry
implies the generation of a spin gap. But:

$i)$ The converse need not be true. We will see that there are spin models
that exhibit \textit{gapped} phases\footnote{The so-called "large-$D$" phases
of the "$\lambda-D$" model to be discussed immediately below.} \ while
retaining the full $Z_{2}{\times} Z_{2}$ symmetry, and:

$ii)$ The mere nonvanishing of (one or more) string order parameters is not
enough to fully determine in which (gapped) phase the system is. It is the
\textit{full set} of order parameters, both string and N\'{e}el, that allows
for a full characterization of the various phases. In particular, the Haldane
phase is fully characterized by the vanishing of all the N\'{e}el order
parameters and by all the three string parameters being nonzero.

We turn now to a different class of models, the so-called "$\lambda-D$" family
of models\footnote{This is the class of models on which the Bologna group is
currently working.}. They are described by the family of Hamiltonians
(parametrized by two real parameters, $\lambda$ and $D$):%
\begin{equation}
\mathcal{H=}%
{\displaystyle\sum\limits_{i=1}^{N}}
\left\{  S_{i}^{x}S_{i+1}^{x}+S_{i}^{y}S_{i+1}^{y}+\lambda S_{i}^{z}%
S_{i+1}^{z}+D\left(  S_{i}^{z}\right)  ^{2}\right\}
\label{ld}
\end{equation}

The "standard" (isotropic) $AFM$ Heisenberg model corresponds of course to
$\lambda=1$ and $D=0$. $\lambda=-1$ (and $D=0$) can be easily
shown\footnote{By performing a rotation of $\pi$ around the $z$-axis on
one of the two sublattices (i.e. on every other site).} to correspond to a
(isotropic) ferromagnetic Heisenberg model. $|\lambda|\neq1$ introduces an
"Ising-like" anisotropy, while $D\neq0$ introduces what is called "single-ion" anisotropy.

The model can be solved exactly for $S=1/2$ \cite{KBI}, but no exact solutions
are available for integer spin. There are obvious asymptotic limits when
either $\lambda$ (resp. $D$) is large and $D$ (resp. $\lambda$) not too large,
so that the "$\lambda$-term" (resp. "$D$-term") can be considered as a
zeroth-order Hamiltonian and the rest as a perturbation:

$i)$ \ $|\lambda|\gg1$. The reference ground state is either a N\'{e}el $AFM$
state ($\lambda>0$) or a ferromagnetic ($\lambda<0$) state.

$ii)$ $|D|\gg1$. \ For $D>0$ (the so-called "large-$D$" phase) the reference
state becomes a planar state with $S_{i}^{z}=0$ for all $i$'s, while for $D<0$
the reference state is a state where $S_{i}^{z}=0$ is excluded, hence a state
where the $S=1$ spins become effectively two-level systems, and a detailed map
of the model into an effective spin-$1/2$ model \cite{DER,Ro} can be
successfully performed. For $|\lambda|\neq1$ and $D\neq0$ the symmetry group
of the Hamiltonian is $O(2){\times} Z_{2}$ (the $Z_{2}$ factor corresponding to
a reflection in the $x-y$ plane: $S_{i}^{z}\to-S_{i}^{z}$).

\begin{figure}[h]
\begin{center}
\includegraphics[height=80mm,keepaspectratio,clip]{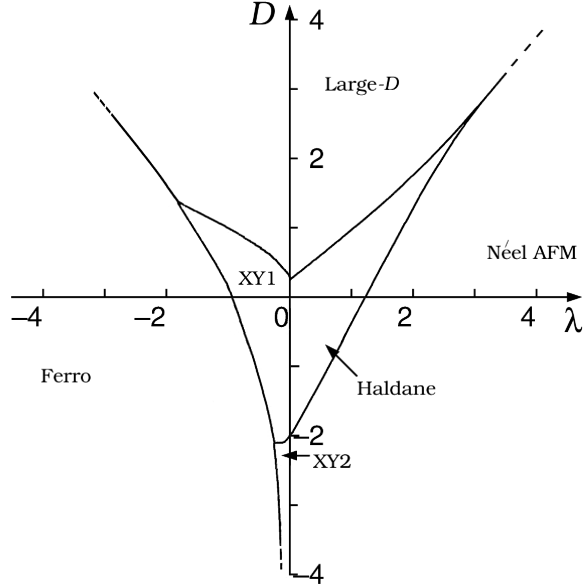}
\caption{Phase diagram of the $\lambda-D$ spin-1 Hamiltonian of Eq. (\ref{ld}).
The indicated regions are explained in the text.}  
\label{Fig1}
\end{center}
\end{figure}

Apart from these limiting cases, the model has been studied
analytically \cite{Schu} as well as numerically \cite{BJK,CHS,GS,So} , and the
corresponding phase diagram is displayed in Fig.\ref{Fig1}.

The various sectors of the phase diagram can be characterized as
follows \cite{CHS,DER,Er}:

$i)$ the \textit{Haldane phase. }The ground state is unique with total
component $S_{tot}^{z}=0$ of the spin. The order parameters are:
$\mathcal{O}_{N}^{\alpha}=0$, $\mathcal{O}_{S}^{\alpha}\neq0,\alpha=x,y,z$.
The (Haldane) gaps are in different spin channels according to the sign of
$D$, but nonzero in any case. The isotropic, Heisenberg point $\lambda
=1,D=0$ is in this phase, and lies on a line separating the two subphases, that
are denoted as $H1$ and $H2$ in the literature \cite{BJK}.

$ii)$ The \textit{N\'{e}el phase}. The ground state is doubly degenerate, and
the order parameters are: $\mathcal{O}_{N}^{\alpha}=\mathcal{O}_{S}^{\alpha
}=0$ for $\alpha=x,y$, but: $\mathcal{O}_{N}^{z},\mathcal{O}_{S}^{z}\neq0$.

$iii)$ The \textit{large-}$D$ \textit{phase}$.$The ground state is unique, it
is \textit{gapped}, but here:$\mathcal{O}_{S}^{\alpha}=\mathcal{O}_{N}%
^{\alpha}=0$ $\forall\alpha$.

$iv)$ The two \textit{$XY$ phases}. These are both \textit{gapless}
phases. They are distinguished by the nature of the low-lying spin excitations
(spin-$1$ in the $XY1$ phase, spin-$2$ in the $XY2$ phase).

$v)$ The \textit{ferromagnetic }phase. The ground state is doubly degenerate,
with maximal magnetization: $S_{tot}^{z}={\pm} N$, and the phase is gapped. \textbf{
}In this case it is the \textit{ferro}magnetic order parameter that is
nonvanishing, and actually \cite{De}: $\mathcal{O}_{ferro}^{z}=1$ (the other
two being zero). Also: $\mathcal{O}_{S}^{z}(j,k)=\left(  -\right)  ^{j-k-1}$, while: $\mathcal{O}_{S}%
^{x,y}=0$.

Anticipating some of the numerical results of Sect.5,
we give below, in Figs.$2$ and $3$, some examples \cite{De} of the behavior of
the various correlators and order parameters as functions of the parameters of the model.

\begin{figure}[h]
\begin{center}
\includegraphics[height=60mm,keepaspectratio,clip]{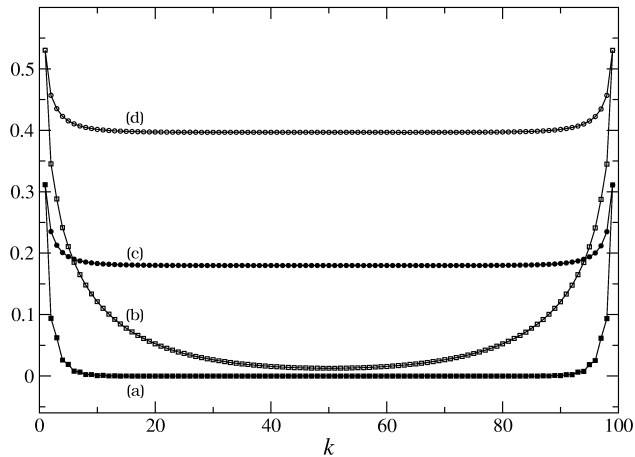}
\caption{Ordinary and string correlation functions in the Haldane phase: 
(a) ${\cal G}^z(k)$,
(b) $\frac{1}{2}(-)^k \langle S_0^+ S_k^- \rangle$,
(c) ${\cal G}_{S}^z(k)$ and 
(d) ${\cal G}_{S}^x(k)$. Selected values
of the parameters are $(D=0.5,\lambda=1)$. Note that with this choice the transverse correlation
length is appreciably larger than the longitudinal one.
The data have been obtained with finite-size $DMRG$
on a chain of $L=100$ spins ($S=1$) with $PBC$ and $M=216$ states
(Sect.5 for details).}  
\label{Fig2}
\end{center}
\end{figure}

\begin{figure}[h]
\begin{center}
\includegraphics[height=60mm,keepaspectratio,clip]{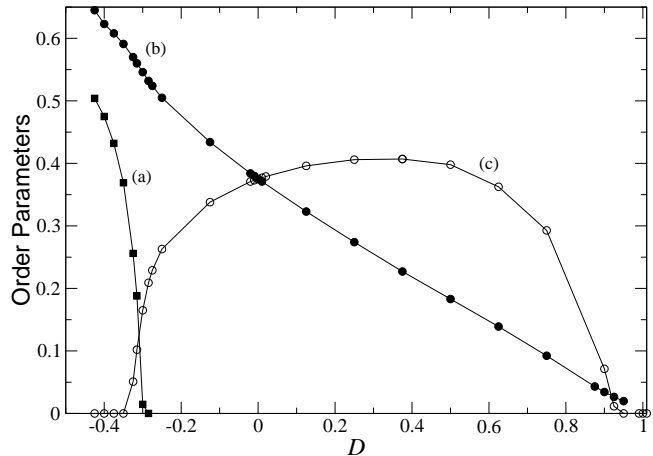}
\caption{Order parameters relevant to the N{\'e}el-Haldane-large $D$ transitions plotted \textit{versus}
the anisotropy coefficient $D$ of Eq. (\ref{ld}) fixing $\lambda=1$: (a) ${\cal O}_N^z$ 
defined in Eq. (\ref{NOPs}); (b) ${\cal O}_S^z$ and (c) ${\cal O}_S^x$ defined in Eq. (\ref{SOPs}).
The asymptotic values are extrapolated using an algebraic best-fit function ${\cal O}_\infty+
C/\vert i-j \vert^\gamma$ on the $DMRG$ data (same choices as in Fig.\ref{Fig2}). 
Near $D \simeq -0.3$ ${\cal O}_N^z$ and ${\cal O}_S^x$ do not vanish in the same point due
to finite-size effects and to a moderate number of $DMRG$ states.}  
\label{Fig3}
\end{center}
\end{figure}

Concerning the nature of the transitions between the various
phases \cite{BJK,CHS,DER}, both the Haldane-large-$D$ and the Haldane-N\'{e}el
transition lines are critical (gapless) lines. The two critical lines merge at
a tricritical point (at $D\simeq\lambda\simeq3$), above which the Haldane
phase disappears and the transition (a large-$D$-N\'{e}el transition,now) is
first-order. The $XY$-ferromagnetic transition is instead a first-order one,
as well as the large-$D$-ferromagnetic transition. Finally, The Haldane-$XY$
transition is considered \cite{CHS} to be of the
Berezinskii-Kosterlitz-Thouless \cite{Ber,KoTh} ($BKT$) \ type, as well as the
$XY$-large-$D$ transition.

The "$\lambda-D$" model has also been studied by $KT$. Applying the same
nonlocal unitary transformation that was discussed previously, they showed
that the transformed Hamiltonian, whose explicit form we will not give here,
is still given in terms of the operators $h_{i}$ (see Eq.$(31$)), and retains
therefore $Z_{2}{\times} Z_{2}$ as the only local symmetry,just as in the case
of the Hamiltonian of Eq. ($16$). Therefore, the same conclusions as before
apply concerning the connection of the nonvanishing of the string order
parameters with the spontaneous breaking of the $Z_{2}{\times} Z_{2}$ symmetry.

In the present paper we will address mainly to the detailed nature of the
Haldane-large-$D$ and Haldane-N\'{e}el critical transition lines. It is known
that the (large distance) critical behavior of one-dimensional quantum systems is well
described by Conformal Field Theory \cite{Ca,DNS,Gi,GNT,KBI} ($CFT$). In the
next Section we will report on a proposal\ of an effective $CFT$ \ for the
"$\lambda-D$" model on the Haldane-large-$D$ critical line. This allows for the prediction of
the operator content of the theory, and hence also for the prediction of the
structure of the conformal tower of excited states above the ground state. To
confirm the predictions, we will report also on extended numerical analyses,
whose details will be reported elsewhere \cite{DeOr}, that fully confirm the
theoretical predictions.

\section{Conformal Field Theory and Effective Actions.}
\label{S_CFT}

Let us  begin by recalling  some basic results and examples of $CFT$ that will be used in the forthcoming analysis of the critical properties of the spin-1 $\lambda-D$ chain. 

It is well known \cite{DNS,GNT} that critical properties of  two-dimensional systems are completely classified by $CFT$'s: 
since in 2$D$ the conformal group is infinite dimensional, the Hilbert space of a conformally invariant theory can be completely understood in terms of  the irreducible representations  of its algebra, the Virasoro algebra. We recall that the latter has an infinite number of generators, denoted with $L_n, \bar{L}_n$ ($n \in \mathbb{Z}$) for its holomorphic and antiholomorphic part respectively, satisfying the commutation relations: 
\be
[L_n,L_m] = (m-n) L_{m+n} + \frac{c}{12} (m^3 -m) \delta_{m+n,0} \label{cr}
\ee
and similarly for the $\bar{L}_n$. The constant $c$ is called the central charge of the algebra or the conformal anomaly. 
Since we are interesting in a comparison between theoretical predictions and numerical data, which are performed on a finite lattice, we will consider a $CFT$ defined on a cylinder with spatial dimension of finite length $L$. 
In this case \cite{Ca,DNS}, the energy and the momentum operator are  represented respectively by:
\bea
H&=& \frac{2\pi}{L} \left( L_0+ \bar{L}_0 - \frac{c}{12}\right)\\
P&=& \frac{2\pi}{L} \left( L_0- \bar{L}_0 \right)
\eea

In order for  $H$ to be bounded from below, we must restrict our attention to highest weight representations of the Virasoro algebra, for which there exists a highest weight (or primary) state $ |\Delta, \bar{\Delta}\rangle$ satisfying:
 \be
L_0 |\Delta,\bar{\Delta}\rangle = \Delta  |\Delta,\bar{\Delta}\rangle   \; , \;  L_n  |\Delta,\bar{\Delta}\rangle =0 \; \;  \mbox{ for } n>0  
 \ee
and  analogous relations with repect to the $ \bar{L}_n $ generators.

Each of these representations is thus identified by the values of the central charge $c$ and of the couple $(\Delta, \bar{\Delta})$ (the conformal dimensions). 
They fix both the energy and the momentum of the  primary state $ |\Delta,\bar{\Delta}\rangle$, according to: 
\bea
E^0_{\Delta,\bar{\Delta}}&=& \frac{2\pi}{L} (\Delta+ \bar{\Delta} -\frac{c}{12})  \\
 P^0_{\Delta,\bar{\Delta}}&=&  \frac{2\pi}{L} (\Delta- \bar{\Delta}) 
\eea
Notice that, in a finite geometry (with $PBC$), 
the vacuum state, corresponding to $\Delta=\bar{\Delta}=0$, has a non zero energy (Casimir effect):
\be
E^0_{vac} = - \frac{\pi c}{6L}
\label{EGSvsL} 
\ee
Also, the two-point correlation function of the operator creating a given  primary state out of the vacuum ($ |\Delta,\bar{\Delta}\rangle = 
{\cal O}_{\Delta,\bar{\Delta}} |0 \rangle $) has an algebraic decay whose critical exponents are determined by the values of the conformal dimensions  $(\Delta, \bar{\Delta})$: one has \cite{DNS,GNT}
\be
\langle {\cal O}_{\Delta,\bar{\Delta}}(z, \bar{z}){\cal O}_{\Delta,\bar{\Delta}}(0,0) \rangle \propto \frac{1}{z^{2\Delta} \bar{z}^{2\bar{\Delta}}} \label{cf}
\ee

Finally, from  the  primary state $|\Delta,\bar{\Delta}\rangle$ one can obtain all excited (or secondary) states  by applying strings of powers of $L_n, \bar{L}_n$ with $n<0$. It is easy to see that, if $m,n <0$, the commutation relations (\ref{cr}) imply :
 $L_0 (L_m)^j  |\Delta,\bar{\Delta}\rangle = (\Delta+ mj)  |\Delta,\bar{\Delta}\rangle$, 
 $\bar{L}_0 (\bar{L}_n)^k  |\Delta,\bar{\Delta}\rangle = (\bar{\Delta} + nk)  |\Delta,\bar{\Delta}\rangle$, so that the secondary states have energies and momenta:
\bea
E^{(r,\bar{r})}_{\Delta,\bar{\Delta}} - E^0_{vac} &=&   \frac{2\pi}{L}(\Delta+ \bar{\Delta} +r+\bar{r}) \label{en} \; , \\
 P^{(r,\bar{r})}_{\Delta,\bar{\Delta}}&=&   \frac{2\pi}{L} (\Delta- \bar{\Delta} +r-\bar{r}) 
\label{EvsdL} 
\eea
with $r,\bar{r} \in {\mathbb N}$ and a degeneracy that can be explicitely calculated for each representation.
It may happen that some of these states have null norms. In this case the true (non-degenerate) Hilbert space of states is obtained after projecting out these null vectors, which therefore do not contribute to the operator content of the corresponding $CFT$. The quantity in brackets in the right hand side of Eq. (\ref{en}) yields the coefficients with which the energy of the corresponding state scales to zero in the thermodynamic limit. It is therefore called "scaling dimension" and will be denoted by $d^{(r,\bar{r})}_{\Delta,\bar{\Delta}}$ in the sequel.
 
Let us examine some examples.
We will consider only unitary theories, corresponding \cite{DNS} to the following 
two sets of values of the central charge  $c$:
\bea
c&=& 1-\frac{6}{p(p+1)}  \; , \; p=3,4,\hdots  \; ; \label{cc}\\
c&\geq &1 \; . 
\eea
The first set of values corresponds to the so called minimal models \cite{DNS,GNT}, whose primary states are of finite number. Their conformal dimensions are given by the formula:
\be
\Delta_{rs} , \bar{\Delta}_{rs} = \frac{[(p+1)r-ps]^2-1}{4p(p+1)} \; \; , \;
1\leq s\leq r\leq p-1 \;   , \; r,s \in \mathbb{Z} \label{rs}
\ee
Theories with $c\geq1$ have  instead an infinite number of primary states. 
 
The simplest case of a $CFT$ corresponds to $c=1/2$ ($p=3$ in Eq.  (\ref{cc})) and describes  the univerality class of the two-dimensional Ising model. According to (\ref{rs}), there are only three primary operators: the identity $\mathbb{I}$ corresponding to the vacuum, $ (\Delta, \bar{\Delta})_{\mathbb{I}} = (0,0)$, the Ising spin $\sigma$ with $ (\Delta, \bar{\Delta})_{\sigma} = (1/16,1/16)$ and the energy density $\varepsilon$ with  $ (\Delta, \bar{\Delta})_{\varepsilon} = (1/2,1/2)$. Notice that the spin-spin correlator $\langle \sigma(x) \sigma(0) \rangle$ decays with a critical exponent $\eta^z = 4 \Delta_{\sigma} = 0.25$. In Table 1 we list the lowest conformal (primary and secondary) states, together with their scaling dimensions and  momenta. As explained in the next section, a  comparison with the numerical data given in the last column will allow us to conclude that the Haldane-N\'{e}el critical transition line is indeed of the Ising type. 

\begin{table}[h]
\caption{\small{Columns 1-4 show the conformal dimensions $(\Delta, \bar{\Delta})$,$(r, \bar{r})$, the scaling dimensions $d^{(r,\bar{r})}_{\Delta,\bar{\Delta}}$ and  the momenta $P^{(r,\bar{r})}_{\Delta,\bar{\Delta}}$ of the lowest conformal states in  the $c=1/2$ minimal model. 
The numerical results in the last column are explained in Sect.5.\newline
Notice that the states with $\Delta=\bar{\Delta}=0$, $(r,\bar{r})= (1,0),(0,1)$ do not appear since they correspond to null vectors. }}
\centering{
\begin{tabular}{ccccc}
\hline
$(\Delta, \bar{\Delta})$ \, , \, $(r, \bar{r})$ & 
$~d^{(r,\bar{r})}_{\Delta,\bar{\Delta}}~$
& $~P^{(r,\bar{r})}_{\Delta,\bar{\Delta}}~$ & $d_{(num)}$ \cr
\hline
$(0,0) \, ; \,(0,0)$  &$ 0$& $0$  & \cr
$(1/16,1/16) \, ;\, (0,0) $ &$ 1/8$ &$ 0  $& $0.1250 \pm 0.0004$ \cr
$(1/2,1/2) \, ; (0,0)$  & $1$ &$ 0$  & $0.962 \pm 0.001$ \cr
$(1/16,1/16) \, ; \,(1,0),(0,1)$  & $9/8$ & $\pm 2\pi/L$ & $1.0959 \pm 0.0008$ \cr
                                  &       &              & $1.100 \pm 0.003$ \cr
$(1/2,1/2) \, ;\,(1,0),(0,1)$  & $2$ &$\pm 2\pi/L $ & $1.87 \pm 0.02$ \cr
                               &     &              & $1.87 \pm 0.02$ \cr
$(0,0) \, ;\, (2,0),(0,2) $ & $2$ &$ \pm 4\pi/L$ & $1.904 \pm 0.004$ & \cr
                            &     &              & $1.86 \pm 0.01$ \cr
\hline
\end{tabular}
}
\label{SDIsing}
\end{table}

We discuss now briefly the $c=1$ case, which exhibits a much richer structure. It corresponds to the field theory of a free compactified bosonic field, i.e. to a  Gaussian model with Lagrangian: 
\begin{equation}
\mathcal{L}=\frac{1}{2}\left[\frac{1}{v}(\partial_{\tau}\Theta)^2+
v(\partial_{x}\Theta)^{2} \right]
\label{o2}
\end{equation}
where $\Theta$ represents an angular variable spanning a circle of a given radius $R$ and 
the constant $v$, which  has the dimension of a  velocity, is called spin velocity. If we assume for $\Theta$, and hence for its dual field $\Phi${\footnote{If we decompose the field $\Theta$ in its holomorphic and antiholomorphic part, $\Theta(z,\bar{z}) = \Theta_h(z) + \Theta_{ah}(\bar{z})$, the dual field is defined as $\Phi = \Theta_h(z) - \Theta_{ah}(\bar{z})$.}, periodic boundary conditions, the Hilbert space of the theory splits into a direct sum of distinct topological sectors labeled by the winding numbers $n,m \in \mathbb{Z}$ of the fields $\Theta$ and $\Phi$ respectively. The primary fields are then vertex operators of the form \cite{DNS,GNT}
\begin{equation}
V_{mn} = \exp{\left(i \sqrt{4 \pi K}n \Phi + i \sqrt{\pi/K} m \Theta\right)}  
\label{ver}
\end{equation}
whose scaling dimensions are   given by
\begin{equation}
d_{mn}=\left( \frac{m^2 }{4 K}+n^2 K \right) \, , \qquad K=\frac{\pi}{R^2} 
\label{gd}
\end{equation}
Notice that the latter depend explicitely on the radius of compactification. Thus we obtain  a different $c=1$ theory for each value of $R$, i.e. of $K$. For example, $K=1$ corresponds (via fermionization \cite{DNS,GNT}) to a 1$D$ model of
free Dirac ($FD$) fermions. The $K=1/2$ point is said to be self-dual ($SD$) since it is invariant under the duality transformation $ \Theta \Leftrightarrow \Phi$,  $ m  \Leftrightarrow n$, while the point $K=2$ corresponds to the $BKT$ 
critical theory. 

We remark also that the energy operator $(\partial \Theta)^2$ has conformal dimension $2$ for any value of $R$ and hence it is always marginal. The effect of adding it to the Lagrangian (\ref{o2}) results only in a change of the coupling constant  in front, which, in turn, can be absorbed into a rescaling of the radius of compactification of $\Theta$.  Thus we generate a continuous line of inequivalent critical $c=1$ theories, corresponding to different values of $K$. 

It is well known \cite{GNT} that the Gaussian model (\ref{o2}) describes the continuum limit of the spin 1/2 XXZ chain with anisotropy parameter $\Delta$,  as long as $-1\leq \Delta \leq 1$. From the exact Bethe-ans{\"a}tz results, one can show 
\cite{GNT} that 
the interesting cases $\Delta =-1, 0,1$  corrrespond to the $SD$, $FD$ and $BKT$ points of the bosonic theory, respectively.
We would like to show now, that the Gaussian model  (\ref{o2}) describes also the critical properties of the spin-1 $\lambda-D$ Hamiltonian (\ref{ld}) on the Haldane-large-D transition line. In doing so, we will also establish a relationship between the coupling constants $D, \lambda$ of the discrete model and those of the continuum theory, namely the spin-wave velocity $v$ and the compactification radius. This will allow us to make quantitative theoretical predictions to be compared, in next section, to the numerical results.

In the spirit of Haldane'a mapping,  we start from a classical solution, which for $D>\la-1$, is a planar state where the unit vectors $\widehat{\Omega}_j(\tau)$ that represent our spins ($\overrightarrow{S}_j \to S\widehat{\Omega}_j(\tau)$ , 
see Sect.2) are \Neel ordered in the $xy$-plane:  $\widehat{\Omega}_j(\tau)=(\cos(\theta_0 + j\pi), \sin(\theta_0 + j\pi),0)$. Hence we make the Haldane-like ans{\"a}tz:
\begin{equation}
\widehat{\Omega}_j(\tau)=(-1)^{j}\hat{n}_j(\tau) \sqrt{1-\frac{l_j^2(\tau)}{S^2}}
+ \hat{z} \; \frac{l_j(\tau)}{S} 
\label{ans}
\end{equation}
where $\hat{n}_j(\tau)= {\rm e}^{{i} \theta_j(\tau)}\in \mathrm{O(2)}_{xy}$, $\hat{z}$ is the unitary vector $(0,0,1)$, and the fluctuation field $l_j$ is supposed to be small. Thus, as for the isotropic case, it is possible to obtain an effective Lagrangian that describes the low-energy physics of the Hamiltonian (\ref{ld}) in the continuum limit.  Carrying out this calculation as explained in Sect.2, one obtains in this case a Gaussian model (\ref{o2}), where now $\Theta = \theta/\sqrt{g}$ and
\begin{equation}
g=\frac{1}{s}\sqrt{2\l(1+D+\la\r)}; \qquad v=s \sqrt{2\l(1+D+\la\r)}
\label{gv}
\end{equation}
In other words, we have a free theory for a  bosonic field $\Theta$, which is  compactified along a circle of radius $1/\sqrt{g}$. Thus, the operator content of the theory can be read from Eq. (\ref{ver}): the list of primary operators is exhausted  by the vertex oprators $V_{mn}$ whose scaling dimensions are given by Eq. (\ref{gd}), with $K=\pi/g$. 

In addition,  the scaling  dimensions (\ref{gd}) fix also the (non universal) critical exponents of the correlation functions. For instance it is easy to see that the transverse spin-spin correlator should decay according to:
\begin{equation}
\langle S^+(0) S^-(x) \rangle\approx \langle {\rm e}^{{i}\theta(0)}  {\rm e}^{-{i}\theta(x)} \rangle
\propto |x|^{-\eta} \mbox{ with } \eta=2d_{10}=g/2\pi \; .
\label{expo}
\end{equation}

\section{The Density Matrix Renormalization Group and Spin Chains.}
\label{S_DMRG}

The code that we have used for density matrix renormalization group ($DMRG$) 
calculations follows rather closely the
algorithms reported in White's seminal papers \cite{Wh92,Wh93}, with the following
points to be mentioned:
\\ \noindent $\bullet$ The superblock geometry was chosen to be $[{\rm B}^s \bullet
\vert {\rm B}^{s'}_{\rm ref} \bullet]$ with $PBC$, where ${\rm B}^{s'}_{\rm ref}$ is the 
(left $\leftrightarrow$ right) reflected of block ${\rm B}^{s'}$ with $s'$ sites.
The rationale for adopting this configuration is that, being effectively on a ring, the two blocks
are always separated by a single site, for which the operators are small matrices that are treated
exactly (no truncation) \cite{Wh93}. In this way we expect a better precision in the correlation functions calculated
fixing one of the two point on these sites and moving the other one along the block. 
Moreover, whenever the system has
an underlying antiferromagnetic structure (typically when a staggered field is switched on), 
this geometry
seems to be the one that preserve it at best, both for even and odd values of $s$.
\\ \noindent $\bullet$ We used the {\it finite-system algorithm} with three iterations. This prescription should
ensure the virtual elimination of the so-called environment error \cite{LF}, which is expected to dominate in the
very first iterations for $L < L^*(m)$ (see below). 
Normally the correlations are computed at the end of the third iteration,
once that the best approximation of the ground state is available.
This has the advantage of using less memory during the finite-size iterations but
requires the storage of all the matrices needed to represent, on the reduced basis of the last step,
the operators entering the correlation functions of interest.
At the moment, disk storage is the ultimate factor that limits the size of the systems
that we are able to treat.
\\ \noindent $\bullet$ We always exploit the conservation of $S_{tot}^{z}$. 
With the exception of the ferromagnetic phase, that we do not address now, 
the ground state(s) is (are) at $S_{tot}^{z}=0$ \cite{BJK}. In order to maximize their accuracy, the
correlations are calculated
targeting only the lowest-energy state within this sector.
However, in order to analyze the energy spectrum,
we had to target also the lowest-energy states in the other sectors
$\vert S_{tot}^{z} \vert=1,2,\dots$ and/or a few excited states within
the $S_{tot}^{z}=0$ sector, depending on the phase of interest. 
On the one hand, this requires a modification of the basic Lanczos method to go beyond
the lowest eigenvalue of the superblock Hamiltonian. On the other hand, once the
$N_{\tt t}$ eigenvalues of interest are found, one can build the block density matrix
as the average (mixture) of the matrices associated with the corresponding $N_{\tt t}$ 
eigenvectors. At present we are not aware of any specific ''recipe'' other than that
of equal weights.

Going back to the modified Lanczos routine, our DMRG code implements
the so-called Thick Restart algorithm of Wu and Simon \cite{WS}.
Once $S_{tot}^{z}$ is fixed, in a given run we want to determine simultaneously the
first $N_{\tt t}$ levels $\vert S_{tot}^{z} ; {\tt b} \rangle$ with {\tt b}=0,1,2,$\dots,N_{\tt t}-1$ 
(the ground state being identified by $(S_{tot}^{z}=0,${\tt b}$=0)$). 
Then, as in the conventional Lanczos scheme,
we have to push the iteration until the norms of the residual vectors
and/or the differences of the energies in consecutive steps
are smaller than prescribed tolerances ($10^{-9}-10^{-12}$ in our calculations). 
The delicate point to keep under control
is that, once the lowest state $\vert S_{tot}^{z};0 \rangle$ is found, if we keep iterating searching for
higher levels the orthogonality of the basis may be lost, just because
the eigenvectors corresponding to these levels tend to overlap again with the vector
$\vert S_{tot}^{z};0 \rangle$. As a result, the procedure is computationally more demanding to
the extent that one has to re-orthogonalize the basis from time to time.
Typically, we have seen that this part takes a 10-20\% of the total time spent
in each call to the Lanczos routine. We have also observed that if this re-orthogonalization
is not performed, one of the undesired effects is that the excited doublets (generally due
to momentum degeneracy) are not correctly computed. More specifically, it seems that while 
the two energy values are nearly the same in the asymmetric stages of the iterations, when the superblock geometry
becomes symmetric ($s=s'$ in the notations of the preceding point) the double degeneracy
is suddenly lost and only one of the two states appears in the numerical spectrum.  

So far for the specific algorithm. Now, the crucial point to consider in accurate $DMRG$ calculations
is the choice of $M$, that is, the number of optimized states. White argued
\cite{Wh93} that the convergence
of the ground state energy is almost exponential in $M$ with a step-like
behaviour, probably related to the successive inclusion of more and more
complete spin sectors. Unfortunately, the effective accuracy gets poorer when we deal with energy
differences and correlation functions, for which little is known about
convergence. It must be told, however, that despite its name the $DMRG$ performs somehow better
for systems with a definite gap rather than for gapless (critical) ones.
We refer to the papers by Andersson, Boman and {\"O}stlund \cite{ABO} and by Legeza and F{\'a}th \cite{LF}
where, for different systems and in terms of different observables, the following common feature
emerges: Even if the quantum system is rigorously critical in the limit $L \to \infty$,
the $DMRG$ truncation introduces a spurious length, $L^*(m)$, which, as expected, diverges as $M$ is
increased. (Our analysis of the accuracy of the energy levels in some selected points of the $\lambda-D$
chain near criticality leads to a similar conclusion \cite{DeOr}). Hence, even if we are technically able
to deal with sizes $L > L^*(m)$ (at a given $M$), as far as criticality is concerned we cannot
rely completely on the $DMRG$ data because the system experiences an effective length which should be
absent in the critical regime. 

Therefore, our strategy can be summarised as follows:
We fix a rather high value of $M$ such that the trustable values of $L$ are sufficiently large
to see the scaling limit of $CFT$, but not too large as compared to $L^*(m)$. In other words, even 
in the study of (supposed) critical systems, we prefer to exploit the computing resources
to include as many $DMRG$ states as possible, and to refine the calculations with finite-size
iterations, rather than trying to take na\"{\i}vely the limit $L \to \infty$.
In addition, to judge whether $M$ is sufficiently large or not we checked the properties
of traslational and reflectional invariance that the correlation functions should have
\footnote{In \cite{DeOr} it is shown that, in general, ${\cal G}_S^{x,y}(j,k)$ behaves
nontrivially under $j \leftrightarrow k$, due to the fact that the ground state
is not necessarily in the $S^z_{tot}=0$ sector.}. To be specific, if ${\cal G}(0,k)$ 
is a certain correlation function computed starting at $j=0$,
we have always increased $M$ (at the expenses of $L$) until the bound 
$\vert {\cal G}(\ell,\ell \pm k)-{\cal G}(0,k) \vert / {\cal G}(0,k) \lesssim 0.05$
was met for $k$ varying from $0$ to $\ell=L/2$, possibly with the exception of the ranges
where ${\cal G}(0,k)$ is below numerical uncertainties ($10^{-6}$, say).

The quality of the numerical analysis of the critical properties depends heavily
on the location of the critical points of interest. 
As far as the transitions from the Haldane phase are concerned, 
it is convenient to fix some representative values of $\lambda$ and let
$D$ vary across the phase boundaries.
This preliminary task of finding $D_{\rm c}(\lambda)$ turns out
to be crucial for subsequent calculations and is divided in two steps.
First, one has to get an approximate idea of the transition points using
a direct extrapolation in $1/L$ of the numerical values of the gaps, computed at
increasing $L$ with a moderate number of $DMRG$ states. Clearly, one may want to explore
a rather large interval of values and so the increments in $D$ will not be particularly
small (0.1, say). Then, the analysis must be refined around the minima of the curves $\Delta E$-vs-$D$
with smaller increments in $D$ and a larger value of $M$. In our problem, the approach that seems
to give better results is standard finite-size scaling ($FSS$) theory \cite{GS,HB} (for instance as compared
to the phenomenological renormalization group).

Once the critical point is located, we take full advantage of the conformal structure
by looking at the finite-size spectrum (see Eqs. (\ref{EGSvsL}) and (\ref{EvsdL}))
of relevant and marginal operators. In practice, we select a number of states
that tend to become degenerate with the ground state and look for straight lines
in the $\Delta E$-vs-$L^{-1}$ plot. Then, from a best fit we expect to have a very
small offset (ideally a zero gap in the thermodynamic limit) and a slope given by the scaling dimension
$d$ multiplied by the velocity prefactor, $v$, which is absent in the field-theoretical formulation
but has to be determined (in terms of the microscopic parameters) in a lattice system.
In the latter case, Eq. (\ref{EGSvsL}) should contain also a term $e_\infty L$, $e_\infty$ being
the energy density of the problem at hand. 
Actually, due to the prefactor $v$, we have to
imagine a self-consistent procedure: Depending on the type of the transition we have in mind
(that is, depending on the central charge $c$), we stick on one or more levels in the
spectrum that have exactly $d=1$. Then the slope of these is nothing but $v$.
Once the velocity is estimated, one uses Eq. (\ref{EGSvsL}) to best fit the product
$cv$ and see whether the value of $c$ and the hypotesis on the universality class are self-consistent
or not. 

To clarify the matter, let us start with the simpler case of the Haldane-N{\'e}el transition, that is
thought to be in the 2$D$ Ising universality class. Fixing $\lambda=0.5$ we find $D_{\rm c}(0.5)=-1.2$, and
the $\beta$-function method \cite{HB} yields $\nu(0.5) = 1.023 \pm 0.009$, as far as the
gap exponent, $\Delta E \propto (D-D_{\rm c})^\nu$ is concerned. Moreover,
we observe the following nontrivial feature of the spectrum:
The massless modes described by the $CFT$ seem to be all and only the levels
within $S_{tot}^z=0$, while those with $S_{tot}^z \neq 0$ mantain a finite energy gap in the
limit of large $L$.
Hence, the reference state for the calculation of $v$ will be the second excited state in $S_{tot}^z=0$,
corresponding to the primary field of conformal dimensions (1/2,1/2).
Using quadratic extrapolations in $1/L$ we get $v=2.44$, and consequently $e_\infty=-2.0011961 \pm 0.0000006$
and $c=0.5008 \pm 0.0008$, thereby confirming the Ising universality class.
The scaling dimensions can be estimated from the slopes of the straight lines in a plot like that
of Fig.\ref{Fig4}. 
In Table \ref{SDIsing} the theoretical values
anticipated in Sect.4 are compared with these numerical estimates. The overall agreement 
is good (7 \% in the worst case). Note that all the marginal operators have nonzero momentum
and so they cannot represent a valid perturbation to the continuum Hamiltonian in as much as
they would break traslational invariance. The absence of marginal operators suggests that
each point of the Haldane-N{\'e}el transition corresponds to the same $c=1/2$ theory and the line
in the phase diagram is ''generated'' by the mapping from the discrete spin model to the
continuum CFT. Repeating the same passages at $\lambda=1$ we get $D_{\rm c}(1)=-0.315$, $\nu(1)=1.003 \pm 0.006$
together with $v=2.65$, $e_\infty=-1.62651$, $c = 0.498 \pm 0.002$, that is, again a $c=1/2$ continuum theory.
\begin{figure}[h]
\begin{center}
\includegraphics[height=80mm,keepaspectratio,clip]{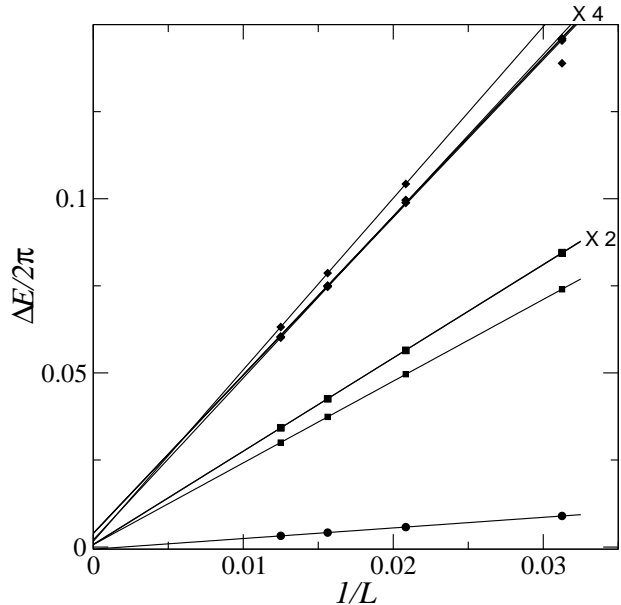}
\caption{Energy differences, divided by $2 \pi$, plotted vs $1/L$ at the Ising transition
$(\lambda=0.5,D=-1.2)$.  
Points represent the numerical values obtained with
multi-target DMRG runs ($M=216$) collecting nine excited states within $S_{tot}^z=0$.
Continuous lines are best-fit whose slopes
are given in Table \ref{SDIsing}, together with the theoretical predictions 
of the scaling dimensions (the labels on the right
indicate the multiplicities, all correctly met).}  
\label{Fig4}
\end{center}
\end{figure}
 
We now pass to an example of numerical investigation of a $c=1$ line of critical points,
namely the transition from the Haldane to the large-$D$ phase.
In the past \cite{DR}, a similarity with the critical fan of the Ashkin-Teller
model has been suggested.
The operator content of this model arises from Ginsparg's orbifold construction \cite{Gi} and
consists of a number of $K$-independent scaling dimensions plus the contributions coming
from the pure Gaussian part (free boson) discussed in the previous Section.  
The fact that we do not observe $K$-independent dimensions (apart from trivial
secondaries of the identity) indicates that the continuum description of our
spin-1 Hamiltonian with $PBC$ at the Haldane-large-$D$ transition should be purely Gaussian rather than 
''orbifold-like''.

In order to support this claim, we try again to match the whole spectrum of the relevant and marginal
operators ($d \leq 2$). The difference with the $c=1/2$ case is that here we have to fix not one but
{\it two} nonuniversal parameters, $v$ and $K$ (see Eq. (\ref{gd})). 
As regards the former, 
the velocity stems from the first and second excited states
in $S_{tot}^z=0$. Note that in choosing these levels
we are assuming, self-consistently, that $K > 1$ so that the two
secondaries of the identity ($d=1$) come first than the primaries with $(m=0,n=\pm 1)$, having 
$d_{0,\pm1}=K$. As far as the Luttinger parameter $K$ is concerned, we have to inspect the
spectrum in other sectors of $S_{tot}^z$ too. In particular, the first excited state
lies in $\vert S_{tot}^z \vert =1$, that corresponds to $m=\pm 1,n=0$ in Eq. (\ref{gd}).
The value of $K$ is obtained from the slope $d_{\pm 1,0}=1/4K$ in a plot similar to
that of Fig.\ref{Fig4}. More generally, in order to check the self-consistency
of the hypotesis $c=1$, we have computed the finite-size spectrum of relevant and marginal
operators in different sectors of $S_{tot}^z$ for a couple of critical points on the Haldane-large-$D$ line
(first two rows of Table \ref{vceiK_HD}). 
Once that $v$ and $K$ are numerically determined, the structure of the Gaussian spectrum
is correctly reproduced (including the multiplicities) and the overall comparison
is satisfactory since, in worst cases, the relative difference does not exceed 3\%
(see plots and tables of Ref. \cite{DER}). 
The agreement with the theoretical predictions of the mapping in the
planar regime is also remarkable.
If we plug the coordinates of the critical points
in the formulae of $g$ and $v$ for the Gaussian model derived above (Eq. (\ref{gv})),
we obtain $v=g=2.07$, $K=\pi/g=1.52$
at $(\lambda=0.5,D=065)$ and $v=g=2.45$, $K=\pi/g=1.28$ at $(\lambda=1,D=0.99)$.

\begin{table}[h]
\caption{Velocity, central charge and ground state energy density for some critical points
on the Haldane-large-$D$ transition line. 
The numbers are the outcome of $DMRG$ calculations with $L=16,20,24,32,48,64$ and $M=405$,
for cases with $K>1$, or $L=16,20,24,28,32,36,40$ and $M=400$, for cases with $K<1$.
The last two columns contain the numerical estimate of the nonuniversal parameter $K$,
(according to the procedures described in the text) and the gap exponent obtained
from the CFT formula $\nu=1/(2-K)$. The error on $e_\infty$ is typically of one unit in the
last digit or better.}
\centering{
\begin{tabular}{cccccc}
\hline
$[\lambda,D_{\rm c}(\lambda)]$ & $v$ & $c$ & $e_\infty$ & $K$ & $\nu$ \\
\hline
$(0.50,0.65)$ & 2.197 $\pm$ 0.004 & 1.008 $\pm$ 0.003 & $-0.908765$ & 1.580 $\pm$ 0.004 & 2.38 \\
$(1.00,0.99)$ & 2.588 $\pm$ 0.006 & 0.997 $\pm$ 0.003 & $-0.859152$   & 1.328 $\pm$ 0.004 & 1.49 \\
$(2.59,2.30)$ & 3.70 $\pm$ 0.04 & 0.99 $\pm$ 0.01 &   $-0.675099$  & 0.85 $\pm$ 0.01 & 0.870 \\
$(3.20,2.90)$ & 4.445 $\pm$ 0.005 & 1.133 $\pm$ 0.006 & $-0.59132$ & 0.526 $\pm$ 0.007 & 0.678 \\
\hline
\end{tabular}
\label{vceiK_HD}
}
\end{table}

Enforced by these quantitative predictions, we try to approach
the multicritical point where the $c=1$ line meets the $c=1/2$ one. Supposedly, the central
charge at this point is $c=3/2$ and it has been proposed \cite{Schu} that the corresponding
CFT is a SU(2)$_2$ Wess-Zumino-Witten-Novikov model. If this was true, the two lines should join
at the point where the effective Gaussian theory has $K=1$ \cite{GNT} ($FD$ point).
Using $\lambda \simeq D$ in the expression of $g$ we find that $K=\pi/g(\lambda)=1$
is satisfied for $\lambda \simeq 2$, while it is believed \cite{CHS} that the multicritical
point lies at $\lambda \gtrsim 3$. We guess that the two lines join at $K < 1$, and in order
to test this conjecture we study two more points: $(\lambda=2.59,D=2.30)$, again on the
$c=1$ line, and $(\lambda=3.20,D=2.90)$
proposed in \cite{CHS} as the multicritical point itself.
Altough the steps are conceptually the same as above, here we encounter two additional
complications. First, due to the closeness (or almost coincidence in the multicritical case)
of the Ising transition, we observe the merging of the two (quasi)critical spectra.
Hence, we have to target more states and separate the ones belonging to $c=1$ from the
ones belonging instead to $c=1/2$. Second, we observe sizeable finite-size corrections
from irrelevant operators. In fact, our analysis shows that we are moving at values
of $K$ smaller than one towards $K=1/2$ where certain irrelevant operators become
marginal. As explained in \cite{DER}, the last two rows of Table \ref{vceiK_HD}
are obtained by extracting $K$ not from the first excited state, but rather
from half the sum of the pair of levels with $m=0,n=\pm 1$ in $S_{tot}^z=0$,
to get rid of finite-size corrections.
As anticipated, moving to
the right of the Haldane-large-$D$ line the value of $K$ keeps on decreasing towards 1/2 ($SD$ point) 
where we argue that this line meets the Haldane-\Neel one and a first order transition starts.
 
We close the section with a few comments on the hidden topological order measured by
string order parameters (Eq. \ref{SOPs}). It is expected that, leaving the Haldane phase,
the $Z_{2}{\times} Z_{2}$ symmetry is partially or totally restored. More precisely, when the
$c=1$ line is crossed, both ${\cal O}_S^z$ and ${\cal O}_S^{x,y}$
vanish. As customary, we can introduce two off-critical exponents that
control the closure of these order parameters. For instance, fixing $\lambda$ and varying
$D$ about $D_{\rm c}(\lambda)$:
\begin{equation}
{\cal O}_S^z \propto (D_{\rm c}-D)^{2 \beta_S^z} \;,\;\; {\cal O}_S^{x} \propto (D_{\rm c}-D)^{2 \beta_S}
\label{betasHD}
\end{equation}
Now, according to $FSS$ arguments (sec. 5.1 of \cite{Gi}),
$\beta_S$ and $\beta_S^z$ are related, via the gap exponent $\nu$, to their counterparts at criticality, that is, 
the scaling dimensions of the operators entering the associated string correlation functions.
These dimensions, in turn, can be extracted from the slopes, $\eta_S$ and $\eta_S^z$, in the log-log plots of 
${\cal O}_S^{x,z}(D=D_{\rm c})$ evaluated at half of the chain. 
Using the relation $2\beta_S=\nu \eta_S$ (and analogously for the $z$ channel) we find the
values reported in Table \ref{FSSexpG} for a couple of critical points already discussed above.
We should observe that the scaling dimensions $\eta_S/2$ and $\eta_S^z/2$ are
{\it not} contained in the $c=1$ spectra cited above. However, we notice also that the numerical
estimates of $\eta_S^z$ are rather close to the values $\frac{2 \; d_{0,\pm1}}{4}=K/2$ and that these
levels actually exist in the effective continuum theory provided that {\it half-intger} values
of $n$ are allowed in Eq. (\ref{gd}). In the XXZ spin-1/2 formulation this is known to
correspond to twisted boundary conditions on the chain. Thus, considering that the
calculations presented here for the spin-1 case are with $PBC$, it's not surprising
that the scaling dimensions associated with ${\cal O}_S^{x,z}$ are absent in the numerical
spectra. Nonetheless, we believe that the closeness to $K/2$ is not accidental and
in Ref. \cite{DER} we speculated about the possibility that the longitudinal string correlation functions
(Eq. (\ref{SOcf}) with $\alpha=z$) acquires, in the continuum limit, the asymptotic form 
\begin{equation}
{\cal G}_S^z(r) \sim \langle \exp{[\mp i \sqrt{\pi K} \Phi(0)]} \exp{[\pm i \sqrt{\pi K} \Phi(r)]} \rangle
\label{SOcfc}
\end{equation}  
so that the lattice string $S_{0}^{z} \exp\left[ i\pi{\sum_{l=1}^{r-1}}
S_{l}^{z} \right]$ is somehow related to the continuum twist operator $\exp{[\pm i \sqrt{\pi K}\Phi(r)]}$.

\begin{table}[h]
\caption{Exponents associated with the vanishing string order parameters at the Gaussian transitions 
taking place at the points indicated in the first column (see text for definitions).  All the numbers
are obtained with $FSS$ on the data at $L=32,48,64,80,100$ and $M=300$.}
\centering{
\begin{tabular}{ccccc}
\hline
$[\lambda,D_{\rm c}(\lambda)]$ & $2\beta_S$ & $2\beta_S^z$ & $\eta_S$ & $\eta_S^z$ \\
\hline
(0.50,0.65) & 0.597 $\pm$ 0.009 & 1.91 $\pm$ 0.02 & 0.251 $\pm$ 0.002 & 0.804 $\pm$ 0.003 \\
(1.00,0.99) & 0.407 $\pm$ 0.002 & 1.10 $\pm$ 0.01 & 0.2733 $\pm$ 0.0006 & 0.741 $\pm$ 0.002 \\
\hline                                    
\end{tabular}
\label{FSSexpG}
}
\end{table}

\section{Conclusions.}

In the present paper we have reviewed, to the best of our knowledge, part of
the status-of the-art concerning Heisenberg spin chains, including biquadratic
interaction terms and various kinds of anisotropies, concentrating on the
r\^{o}le of hidden symmetries in the various families of spin models. We have
discussed how the inclusion of anisotropy terms can drive the "standard"
Heisenberg chain away from the Haldane phase and how hidden symmetries (and
their spontaneous breaking) are of great help in classifying the "massive"
(gapfull) phases of the model. The location of the critical lines of the model
has been accurately obtained numerically, confirming and extending earlier
predictions \cite{CHS}.

The combined use proposed here of analytical $\left( CFT\right)  $ and
numerical $\left( DMRG\right)  $ techniques to investigate the critical
properties of the models has proved to be a rather successful strategy to
clarify the nature and structure of the critical phases of the models.
Numerical simulation techniques (Monte Carlo and $DMRG$, to quote only the
most known ones) are of more and more frequent and extended use in almost all
branches of Theoretical Physics. \ A blind use of them can however be more
dangerous than helpful in understanding the physical properties of the systems
for whose study they are employed. We believe instead that an "educated" use
of numerical techniques in support of analytical approaches, as described
here, can result in a powerful synergy that can be of great help in
understanding the physics of many problems in Theoretical Physics.


\section*{Acknowledgments}

The authors would like to thank F.Ortolani
and M.Roncaglia for useful discussions and for a critical reading of the
manuscript. One of us (G.M.) would like to thank the organizer of the
$XIII-th$ Conference on "\textit{Symmetries in Physics}", Prof. Bruno Gruber,
for inviting him to take part in the Bregenz Conference, where the main
content of the present paper was presented.

\end{document}